\documentclass[12pt]{iopart}
\usepackage{iopams}  
\usepackage{xy} \xyoption{all}

\newenvironment{pagi}[1]{\begin{minipage}[t]{#1em}}{\end{minipage}}

\newcommand{\C}{\mathbb{C}}

\newcommand{\hache}{\mathbb{H}}

\newcommand{\R}{\mathbb{R}}

\newcommand{\pint}[2]{\left\langle #1|#2 \right\rangle}

\newcommand{\vcg}[1]{\mbox{$\mathbf{#1}$}}
\newcommand{\vcgs}[1]{\mbox{\scriptsize $\mathbf{#1}$}}

\newcommand{\id}[1]{\mbox{\rm Id}_{#1}}
\newcommand{\imb}{\mbox{\it \i}}
\newcommand{\imbb}{\textbf{\textit{\i}}}
\newcommand{\imbbs}{\scriptsize\textbf{\textit{\i}}}

\newcommand{\drc}[1]{\left|#1\right\rangle}
\begin{document}

\title[Quregisters, symmetry groups and Clifford algebras]{Quregisters, symmetry groups and Clifford algebras}

\author{Dalia Cervantes, Guillermo Morales-Luna}

\address{Computer Science Department, CINVESTAV-IPN, Mexico City, Mexico}
\ead{gmorales@cs.cinvestav.mx}
\vspace{10pt}
\begin{indented}
\item[]November 2015
\end{indented}

\begin{abstract}
%      The Clifford algebra over the three-dimensional real linear space includes its linear structure  and its exterior algebra, the subspaces spanned by multivectors of the same degree determine a gradation of the Clifford algebra. Through these geometric notions, 
Natural one-to-one and two-to-one homomorphisms from $\mbox{SO}(3)$ into $\mbox{SU}(2)$ are built conventionally, and the collection of qubits, is identified with a subgroup of $\mbox{SU}(2)$. This construction is suitable to be extended to corresponding tensor powers. The notions of qubits, quregisters and qugates are translated into the language of symmetry groups. The corresponding elements to entangled states in the tensor product of Hilbert spaces reflect entanglement properties as well, and in this way a notion of entanglement is realised in the tensor product of symmetry groups. 
\end{abstract}

\section{Introduction}

The quantum computations may be realised as geometric transformations of the three-dimensional real space~\cite{Hav01,lomonaco02}. Several approaches have been published within this geometric approach~\cite{Caf10,Do02}. For instance, in~\cite{14Johansson} a relation is reviewed between local unitary symmetries and entanglement invariants of quregisters from the algebraic varieties point of view, and in~\cite{15Feng} a construction is provided via symmetric matrices for pure quantum states where their reductions are maximally mixed. Through the well known identification of the symmetry group $\mbox{\ SU}(2)$ with the unit sphere of the two-dimensional complex Hilbert space, namely, the collection of qubits, the sphere is provided with a group structure. The qu-registers, as tensor products of qubits, in the sense of Hilbert spaces, correspond then to group tensor products. We pose the problem to compute effectively this correspondence and to transfer the notion of entangled states into  the corresponding elements of group tensor products.

\section{The Clifford algebra of three-dimensional space}

For $\R^3$ with the quadratic form ${\bf x}\mapsto \|{\bf x}\|^2 = |x_0|^2 + |x_1|^2 + |x_2|^2$ the Clifford algebra is 
$$\mbox{Cl}(\R^3,\|\cdot\|^2) = \left(\bigoplus_{n\geq 0}\left(\R^3\right)^{\otimes n}\right)/\left\langle({\bf x}{\bf x} - \|{\bf x}\|^2{\bf 1})_{{\bf x}\in\R^3}\right\rangle.$$
Besides
$$\forall{\bf x},{\bf y}\in\R^3:\ \pint{{\bf x}}{{\bf y}} = \frac{1}{2}\left({\bf x}\,{\bf y} + {\bf y}\,{\bf x}\right)\ \ \ ,\ \ \ {\bf x}\land {\bf y} = \frac{1}{2}\left({\bf x}\,{\bf y} - {\bf y}\,{\bf x}\right)$$
where $\land$ denotes the usual exterior product. For any $k\in\{0,1,2,3\}$, let $\bigwedge^k\R^3$ be the linear space spanned by the $k$-vectors. The exterior algebra is $\bigwedge\R^3 =\bigoplus_{k=0}^3 \bigwedge^k\R^3$, hence it is a real $2^3$-dimensional linear space. If $\{{\bf s}_1,{\bf s}_2,{\bf s}_3\}$ is an orthonormal basis of the space of vectors $\bigwedge^1\R^3\approx\R^3$ then:
${\bf s}_i^2 = {\bf 1}\ \&\ \left(i\not=j\ \Rightarrow\ {\bf s}_i{\bf s}_j = -{\bf s}_j{\bf s}_i\right).$
It is naturally identified ${\bf s}_i\leftrightarrow\sigma_i$, for the Pauli matrices
$$\sigma_1 = \left(\begin{array}{rr}
 0 & 1 \\
 1 & 0
\end{array}\right)\ ,\ 
\sigma_2 = \left(\begin{array}{rr}
 0 & -\imb \\
 \imb & 0
\end{array}\right)\ ,\ 
\sigma_3 = \left(\begin{array}{rr}
 1 & 0 \\
 0 & -1
\end{array}\right)\in U(\hache)$$
where $\hache=\C^2$ is the complex 2-dimensional Hilbert space, and $\imb=\sqrt{-1}$. Since $\sigma_1\sigma_2\sigma_3 = \imb\ \id{3}$, it can be defined $\imbb={\bf s}_1{\bf s}_2{\bf s}_3$.
In the subspace of bivectors, let 
${\bf t}_1 = {\bf s}_2 {\bf s}_3$,  
${\bf t}_2 = {\bf s}_3 {\bf s}_1$ and 
${\bf t}_3 = {\bf s}_1 {\bf s}_2$. 
Then, if ${\bf t}_i={\bf s}_j{\bf s}_k$, 
${\bf t}_i^2 =  -{\bf 1}$
and, 
${\bf t}_2{\bf t}_1 = {\bf t}_3$, ${\bf t}_1{\bf t}_3  = {\bf t}_2 $, ${\bf t}_3{\bf t}_2 = {\bf t}_1$,
namely, $\{{\bf t}_1,{\bf t}_2,{\bf t}_3\}$ generates a subalgebra in $\mbox{Cl}(\R^3,\|\cdot\|^2)$ that is isomorphic to Hamilton's quaternions. This algebra is called the {\em even subalgebra} of $\mbox{Cl}(\R^3,\|\cdot\|^2)$. Let ${\cal C}_3=\mbox{Cl}(\R^3,\|\cdot\|^2)$ and ${\cal C}_3^+$ be its even subalgebra.

\section{Symmetry groups of small dimension linear spaces}

For any $\theta\in[-\pi,+\pi]$, and unitary vector ${\bf r}\in\R^3$, let
$$\cos(\theta){\bf 1}+\sin(\theta)\,\imbb\, {\bf r} = \sum_{j\geq 0}\frac{1}{j!}(\imbb {\bf r}\,\theta)^j = e^{\imbbs\, {\bf r}\theta}.$$
$e^{\imbbs\, {\bf r}\theta}$ is an unitary vector in the even subalgebra ${\cal C}_3^+$, thus, it is properly an unitary quaternion, 
$e^{\imbbs\, {\bf r}\theta} = a_0{\bf 1} + a_1{\bf t}_1 + a_2{\bf t}_2 + a_3{\bf t}_3$
with $|a_0|^2 + |a_1|^2 + |a_2|^2 + |a_3|^2 = 1$. 

Through the correspondence
$$\Phi:e^{\imbbs\, {\bf r}\theta}\ \mapsto\ U=\left(\begin{array}{rr}
 a_0 + \imb\, a_3 & a_1 + \imb\, a_2 \\
-a_1 + \imb\, a_2 & a_0 - \imb\, a_3
\end{array}\right) = a_0\id{\hache} + \imb\, a_2\sigma_1 + \imb\, a_1\sigma_2 + \imb\, a_3\sigma_3$$
$\mbox{SO}(3)$ is identified with $\mbox{SU}(2)$, the group of unitary transforms in $\hache$ which preserve the orientation: $e^{-\imbbs\, {\bf r}\frac{\theta}{2}}$ is a rotation of angle $\theta$ with ${\bf r}$ as axis of rotation.

A second identification of $\mbox{SO}(3)$ with $\mbox{SU}(2)$ is obtained by the composition of the following maps:
\begin{equation}
(\theta,{\bf r}) \stackrel{\Psi_0}{\mapsto} {\bf c} = \left[\begin{array}{c}
 c_0 \\ c_1 
\end{array}\right] =  \left[\begin{array}{c}
\cos(\theta) - \imb\, r_3\sin(\theta) \\
r_1\sin(\theta) + \imb\, r_2\sin(\theta) \\
\end{array}\right] %\\
 \stackrel{\Psi_1}{\mapsto} C = \left[\begin{array}{rr}
 c_0 & -\overline{c_1} \\
 c_1 & \overline{c_0}
 \end{array}\right]
 \label{eq.psi1}
\end{equation}
Any rotation in $\mbox{SO}(3)$ is seen as a {\em spinor}, and through the map $\Psi_0$ it corresponds to a qubit ${\bf c}\in S_1 = \{(c_0,c_1)\in\hache|\ |c_0|^2 + |c_1|^2 = 1\}$. Via the map $\Psi_1$, any qubit is identified with an element of $\mbox{SU}(2)$. Conversely, if $C = \left[\begin{array}{rr}
 c_0 & d_0 \\
 c_1 & d_1
 \end{array}\right]\in\mbox{SU}(2)$ then 
 $c_0d_1-c_1d_0=1.$
 By assuming $(c_0,c_1)\in S_1$, the solutions of this equation are the points $(d_0,d_1)\in\hache$ in the straight line passing through $(-\overline{c_1},\overline{c_0})$ parallel to the straight-line orthogonal to $(c_0,-c_1)$. Since $(c_0,-c_1)\in S_1$, this line is tangent to $S_1$ at this point. The solution line is parameterised thus as
 $D(d) = (-\overline{c_1},\overline{c_0}) + d(1,-\overline{c_0^{-1}c_1})$, with $d\in\C.$
 Also $(-\overline{c_1},\overline{c_0})\in S_1$, and the only solution $(d_0,d_1)$ of  $c_0d_1-c_1d_0=1$ in $S_1$ is $(d_0,d_1) =  (-\overline{c_1},\overline{c_0})$. Thus $C=\Psi_1(c_0,c_1)$. Hence, $\Psi_1$ is a bijection $S_1\to\mbox{SU}(2)$. The operation in the group $\mbox{SU}(2)$ translated into $S_1$ is
 $$\star_1:S_1\times S_1\to S_1\ \ ,\ \ 
 \left( \left[\begin{array}{c} c_{00} \\ c_{10} \end{array}\right] , 
 \left[\begin{array}{c} c_{01} \\ c_{11} \end{array}\right]\right)
 \mapsto
  \left[\begin{array}{c} c_{00} \\ c_{10} \end{array}\right] \star_1 
 \left[\begin{array}{c} c_{01} \\ c_{11} \end{array}\right] =
 \left[\begin{array}{c} c_{00}c_{01} -\overline{c_{10}} c_{11} \\ c_{10}c_{01} + \overline{c_{00}} c_{11} \end{array}\right],$$
hence $(S_1,\star_1,\left[\begin{array}{c} 1 \\ 0 \end{array}\right])$ is a group.

\section{Tensor products}

\subsection{Tensor product of Hilbert spaces}

For two complex Hilbert spaces $H_0,H_1$ their tensor product  $H_0\otimes H_1 = \mbox{\rm Lin}(H_1^*,H_0)$ satisfies a proper {\em Universal Property}:
\begin{quote}
 For any space $H_2$ and any continuous bilinear map $\phi:H_0\times H_1\to H_2$ there is a unique bounded linear map $\psi:H_0\otimes H_1\to H_2$ such that $\phi=\psi\circ\iota$, where $\iota:H_0\times H_1\to H_0\otimes H_1$, $({\bf x},{\bf y})\mapsto{\bf x}\otimes{\bf y}$, is the canonical embedding.
\end{quote}
$H_0\otimes H_1$ has as basis the Kroenecker product of corresponding basis in $H_0$ and $H_1$.

\subsection{Tensor product of groups}

Let $V$ be a vector space and $\mbox{\rm GL}(V)$ its group of linear automorphisms. A {\em group representation} of a group $G$ is a group homomorphism $\pi:G\to\mbox{\rm GL}(V)$. 
If $V$ is a complex finite-dimensional linear space, of dimension $n$, then $\mbox{\rm GL}(V)$ is isomorphic to $\mbox{\rm GL}(n,\C)$ and by composing the representation $\pi$ with this isomorphism it is obtained a group homomorphism $G\to\mbox{\rm GL}(n,\C)$. 

Suppose that $\pi_0:G_0\to\mbox{\rm GL}(n_0,\C)$ and $\pi_1:G_1\to\mbox{\rm GL}(n_1,\C)$ are two finite dimensional representations of the groups $G_0$ and $G_1$.  Let $\left(\drc{i}\right)_{i=0}^{n_0-1}$ and $\left(\drc{j}\right)_{j=0}^{n_1-1}$ be basis of $\C^{n_0}$ and $\C^{n_1}$ respectively. Then the tensor product $\C^{n_0}\otimes\C^{n_1} = \C^{n_0n_1}$ has as basis $\left(\drc{i}\otimes\drc{j}\right)_{(i,j)\in\{0,\ldots,n_0\}\times\{0,\ldots,n_1\}}$.
For any pair $(g_0,g_1)\in G_0\times G_1$ let $P_{(g_0,g_1)}:\C^{n_0n_1}\to\C^{n_0n_1}$ be the linear map such that
$$\forall (i,j)\in\{0,\ldots,n_0\}\times\{0,\ldots,n_1\}:\ P_{(g_0,g_1)}(\drc{i}\otimes\drc{j}) = \pi_0(g_0)\,(\drc{i})\otimes \pi_1(g_1)\,(\drc{j}).$$
The set of vectors $\left(P_{(g_0,g_1)}(\drc{i}\otimes\drc{j})\right)_{(i,j)\in\{0,\ldots,n_0\}\times\{0,\ldots,n_1\}}$ is linearly independent, hence $P_{(g_0,g_1)}\in\mbox{\rm GL}(n_0n_1,\C)$. Consequently, the operator 
$$\pi_0\otimes\pi_1:G_0\times G_1\to\mbox{\rm GL}(n_0n_1,\C)\ ,\ (g_0,g_1)\mapsto (\pi_0\otimes\pi_1)(g_0,g_1) = P_{(g_0,g_1)}$$
is an embedding $G_0\times G_1\to\mbox{\rm GL}(n_0n_1,\C)$.

Let $F_{01} = (\pi_0\otimes\pi_1)(G_0\times G_1)$ be the image of $G_0\times G_1$ under the embedding $\pi_0\otimes\pi_1$ and $\left\langle F_{01}\right\rangle$ be the subgroup of $\mbox{\rm GL}(n_0n_1,\C)$ spanned by $F_{01}$. Then the group tensor product $G_0\otimes G_1$ is represented by the group $\left\langle F_{01}\right\rangle<\mbox{\rm GL}(n_0n_1,\C)$.
The group tensor product satisfies the following {\em Universal Property}:
\begin{quote}
 For any group $J$ and any bilinear map $\phi:G_0\times G_1\to J$ there is a unique homomorphism $\psi:G_0\otimes G_1\to J$ such that $\phi=\psi\circ\iota$.
\end{quote}
Thus the group tensor product is associative. 

For any group $G$ with a finite-dimensional representation, its tensor powers are defined recursively:
$$G^{\otimes 2}= G\otimes G\ \ \&\ \ \forall n> 2:\ G^{\otimes n}= G^{\otimes (n-1)}\otimes G.$$

\section{Embedding of $S_1^{\otimes n}$ into $\mbox{SU}(2)^{\otimes n}$}

\subsection{General embedding}

Let $\hache_1=\hache$ and $\forall n>1$: $\hache_n = \hache_{n-1}\otimes\hache$. 
Clearly, $\dim \hache_n = 2^n$. The unit sphere $S_{2^n-1}$ of $\hache_n$ is the set of $n$-quregisters.

The map $\Psi_1$ defined at relation~(\ref{eq.psi1}) is a representation of the group $(S_1,\star_1,\left[\begin{array}{c} 1 \\ 0 \end{array}\right])$, isomorphic to $\mbox{\rm SU}(2)<\mbox{\rm GL}(2,\C)$. Then, the group tensor power $S_1^{\otimes n}$ is isomorphic to the group $\mbox{SU}(2)^{\otimes n} = \left\langle F_1^{\otimes n}\right\rangle<\mbox{\rm GL}(2^n,\C)$ spanned by $F_1^{\otimes n} = \left(\Psi_1(S_1)\right)^{\otimes n}$ in $\mbox{\rm GL}(2^n,\C)$.
Namely, the elements in $\mbox{SU}(2)^{\otimes n}$ can be written as finite length words of the form $T_0^{\varepsilon_0}\cdots T_{k-1}^{\varepsilon_{k-1}}$, with $T_i\in F_1^{\otimes n}$, $\varepsilon_i\in\{-1,+1\}$ and concatenation assumed as map composition, being two such words equivalent if they represent the same element in $\mbox{\rm GL}(2^n,\C)$. In this context, the length of the minimal length word representing an element $T\in\mbox{SU}(2)^{\otimes n}$ can be seen as a degree of separability of the map $T:\hache_n\to\hache_n$.

Let $G$ be any of $\mbox{SO}(3)$ or $\mbox{SU}(2)$ and let $\phi$ be an embedding $S_1\to G$.
By acting component-wise, there is a natural embedding $\phi^n:S_1^n\to G^n$. Then there are maps $\alpha_n$, due to the Universal Property of the Hilbert Space Tensor Product, and $\beta_n$, due to the Universal Property of the Group Tensor Product, that make the following diagram commutative:
\begin{equation}
\xymatrix{
& S_1^n \ar[r]^{\phi^n} \ar[ld]_{\iota_n} & G^n \ar[rd]^{\iota_n} & \\
S_1^{\otimes n} \ar@{-->}[rru]^{\alpha_n}  & & &  G^{\otimes n}\ar@{-->}[llu]_{\beta_n}
}\label{eq.dg01}
\end{equation}
Again by the Universal Properties there are two maps $\psi_n$ and $\omega_n$ commuting the following diagram:
\begin{equation}
\xymatrix{
& S_1^n \ar[r]^{\phi^n} \ar[ld]_{\iota_n} & G^n \ar[rd]^{\iota_n} & \\
S_1^{\otimes n} \ar[rru]^{\alpha_n} \ar@/_/@{-->}[rrr]_{\psi_n} & & &  G^{\otimes n}\ar[llu]_{\beta_n} \ar@/_/@{-->}[lll]_{\omega_n}
}\label{eq.dg02}
\end{equation}
Indeed if the elements in the group $G$ are represented by matrices, then the map $\psi_n$ is transforming vector tensor product into matrix tensor products. Hence, $G^{\otimes n}$ can be identified with $S_1^{\otimes n}$.

Besides, there exists a bijection $\psi_n:S_{2^n-1}\to \mbox{SU}(2)^{\otimes n}$, where $S_{2^n-1}$ is the unit sphere of $\hache_n$ and through this bijection, the collection $S_{2^n-1}$ of $n$-quregisters is realised as the $n$-fold group tensor product of $S_1\approx\mbox{SU}(2)$.

\subsection{Examples of the embedding for lower dimensions}

Let us consider the following pairs of index and sign matrices:
$$\begin{array}{c|c}
n=1 & n=2 \\ \hline
 \begin{array}{cc}
 I_1=\left[\begin{array}{cc}
 0 & 1 \\
 1 & 0 %\\
\end{array}\right] &
\Sigma_1 = \left[\begin{array}{rr}
 1 & -1 \\
 1 & 1 %\\
\end{array}\right]
\end{array} &
 \begin{array}{cc}
 I_2=\left[\begin{array}{cccc}
 0 & 1 & 2 & 3 \\
 1 & 0 & 3 & 2 \\
 2 & 3 & 0 & 1 \\
 3 & 2 & 1 & 0 %\\
\end{array}\right] &
\Sigma_2 = \left[\begin{array}{rrrr}
 1 & -1 & -1 & 1 \\
 1 & 1 & -1 & -1 \\
 1 & -1 & 1 & -1 \\
 1 & 1 & 1 & 1 %\\
\end{array}\right]
\end{array}
\end{array}$$
$$\begin{array}{c}
n=3 \\ \hline
 \begin{array}{cc}
 I_3=\left[\begin{array}{cccccccc}
 0 & 1 & 2 & 3 & 4 & 5 & 6 & 7 \\
 1 & 0 & 3 & 2 & 5 & 4 & 7 & 6 \\
 2 & 3 & 0 & 1 & 6 & 7 & 4 & 5 \\
 3 & 2 & 1 & 0 & 7 & 6 & 5 & 4 \\
 4 & 5 & 6 & 7 & 0 & 1 & 2 & 3 \\
 5 & 4 & 7 & 6 & 1 & 0 & 3 & 2 \\
 6 & 7 & 4 & 5 & 2 & 3 & 0 & 1 \\
 7 & 6 & 5 & 4 & 3 & 2 & 1 & 0 %\\
\end{array}\right] &
\Sigma_3 = \left[\begin{array}{rrrrrrrr}
 1 & -1 & -1 & 1 & -1 & 1 & 1 & -1 \\
 1 & 1 & -1 & -1 & -1 & -1 & 1 & 1 \\
 1 & -1 & 1 & -1 & -1 & 1 & -1 & 1 \\
 1 & 1 & 1 & 1 & -1 & -1 & -1 & -1 \\
 1 & -1 & -1 & 1 & 1 & 1 & -1 & -1 \\
 1 & 1 & -1 & -1 & 1 & -1 & -1 & -1 \\
 1 & -1 & 1 & -1 & 1 & -1 & 1 & -1 \\
 1 & 1 & 1 & 1 & 1 & 1 & 1 & 1 %\\
\end{array}\right]
\end{array}
\end{array}$$
Let $\rho_n = \mbox{\rm exp}\left(\imb\ \frac{2\pi}{2^n}\right) = \mbox{\rm exp}\left(\imb\ \frac{\pi}{2^{n-1}}\right)$ be the $2^n$-th primitive root of unit. 

Then, the corresponding bijections $\psi_n:S_{2^n-1}\to\mbox{SU}(2)^{\otimes n}$ are
$$\begin{array}{lclcrcl}
\psi_1:S_1 &\to& \mbox{SU}(2) &,& [x_i]_{i=0}^{2^1-1} &\mapsto& \left[\sigma_{1ij}\rho_1^j\,x_{I_{1ij}}\right]_{i,j=0}^{2^1-1} \vspace{1ex} \\ 
\psi_2:S_1^{\otimes 2} &\to& \mbox{SU}(2)^{\otimes 2} &,& [x_i]_{i=0}^{2^2-1} &\mapsto& \left[\sigma_{2ij}\rho_2^j\,x_{I_{2ij}}\right]_{i,j=0}^{2^2-1} \vspace{1ex} \\ 
\psi_3:S_1^{\otimes 3} &\to& \mbox{SU}(2)^{\otimes 3} &,& [x_i]_{i=0}^{2^3-1} &\mapsto& \left[\sigma_{3ij}\rho_3^j\,x_{I_{3ij}}\right]_{i,j=0}^{2^3-1} %\\
\end{array}$$
where $I_{kij}$ and $\sigma_{kij}$ are that $ij$-th entries of matrices $I_k$ and $\Sigma_k$, respectively.

It can be seen in a direct way that each identification $\psi_n$ coincides with the linear operator tensor product when restricted to separable states.

\section{Maximally non-separable elements}

Let  $Q=\{0,1\}$ be the set of classical bits and let $\{\drc{0},\drc{1}\}$ denote the canonical basis of $\hache_1$.

For any $n\geq 1$ and any $\vcg{\varepsilon} = \varepsilon_{n-1}\cdots\varepsilon_1\varepsilon_0\in Q^n$ let
$\drc{\vcg{\varepsilon}} = \drc{\varepsilon_{n-1}}\otimes \cdots\otimes\drc{\varepsilon_1}\otimes\drc{\varepsilon_0}.$
Then $\left(\drc{\vcg{\varepsilon}}\right)_{\vcgs{\varepsilon}\in Q^n}$ is the canonical basis of $\hache_n$.

For any $n\geq 2$ and any $\vcg{\varepsilon} = \varepsilon_{n-1}\cdots\varepsilon_1\varepsilon_0\in Q^n$ let
${\bf b}_{\vcgs{\varepsilon}} = \frac{1}{\sqrt{2}}\left(\drc{0\varepsilon_{n-2}\cdots\varepsilon_1\varepsilon_0} + (-1)^{\varepsilon_{n-1}}\drc{1\overline{\varepsilon_{n-2}}\cdots\overline{\varepsilon_1}\,\overline{\varepsilon_0}}\right).$
Then $\left({\bf b}_{\vcgs{\varepsilon}}\right)_{\vcgs{\varepsilon}\in Q^n}$ is the Bell basis of $\hache_n$, and it consists of maximally entangled states.

The corresponding image of linear automoprhisms $$\left(\psi_n\left({\bf b}_{\vcgs{\varepsilon}}\right)\right)_{\vcgs{\varepsilon}\in Q^n}\subset\mbox{SU}(2)^{\otimes n}<\mbox{\rm GL}(2^n,\C)$$ is a collection of maximally entangled elements in $\mbox{SU}(2)^{\otimes n}$.

In this way, the notion of entanglement is transported directly into the context of geometric transformations.   

Since the even algebra ${\cal C}_3^+$ of the Clifford algebra $\mbox{Cl}(\R^3,\|\cdot\|^2)$ is isomorphic to Hamilton's quaternions, the calculation procedures given in~\cite{Gira04,Gira07} may be used within the context of the current paper.

\section{Conclusion}

The main purpose of the current paper is the translation of the most basic notions of Quantum Computing into symmetry groups. Through this translation, qubits become orientation preserving unitary transforms, and quregisters become tensor products of orientation preserving unitary transforms. Thus, the collection of qubits acquire a group structure and the collection of quregisters, which are tensor products of qubits, acquire as well a group structure which is, in a consistent
way, the group tensor product of the corresponding quit group structure. By the way, using group representations, the involved group tensor products can efficiently be represented, e.g., in a manageable way.

\section*{References}

%\bibliographystyle{plain}
%\bibliographystyle{jphysicsB}
%\bibliography{qufest}

\end{document}